\documentclass[final, aps, pra, reprint, citeautoscript, superscriptaddress, nobalancelastpage]{revtex4-1}
\usepackage[T1]{fontenc}
\usepackage[utf8]{inputenc}
\usepackage[english]{babel}

\usepackage{lmodern}
\usepackage{textcomp}
\usepackage{amsmath}
\usepackage{amssymb}
\usepackage{bm}

\usepackage[colorlinks, urlcolor=blue, citecolor=blue, linkcolor=blue, pdfstartview=FitH]{hyperref}
\usepackage[all]{hypcap}
\usepackage[dvipsnames]{xcolor}
\usepackage{subfigure}
\usepackage[pdftex]{graphicx}

\usepackage[protrusion=true, kerning=true, expansion=true]{microtype}

\usepackage{color}

\usepackage[normalem]{ulem}

\def\affiSOLAB{Spin\ Optics\ Laboratory, Saint~Petersburg\ State\ University, 198504\ St.~Petersburg, Russia}
\def\affiIOFFE{Ioffe\ Institute, Russian\ Academy\ of\ Sciences, 194021\ St.~Petersburg, Russia}
\def\affiPhysDep{Department\ of\ Physics, Saint~Petersburg\ State\ University, 198504\ St.~Petersburg, Russia}
\def\affiPhotonics{Department\ of\ Photonics, Saint~Petersburg\ State\ University, 198504 St.~Petersburg, Russia}

\begin{document}

\author{A.~A.\ Fomin}
\author{M.~Yu.\ Petrov}
\author{G.~G.\ Kozlov}
\affiliation{\affiSOLAB}
\author{M.~M.~Glazov}
\affiliation{\affiIOFFE}
\affiliation{\affiSOLAB}
\author{I.~I.\ Ryzhov}
\affiliation{\affiPhotonics}
\affiliation{\affiSOLAB}
\author{M.~V.\ Balabas}
\affiliation{\affiPhysDep}
\author{V.~S.\ Zapasskii}
\affiliation{\affiSOLAB}

\title{Spin-alignment noise in atomic vapor}
\date{\today}
\begin{abstract}
In the conventional spin noise spectroscopy, the probe laser light monitors fluctuations of the \emph{spin orientation} of a paramagnet revealed as fluctuations of its gyrotropy, i.e., circular birefringence. 
For spins larger than $1/2$, there exists spin arrangement of a higher order---the \emph{spin alignment}---which also exhibits spontaneous fluctuations. 
We show theoretically and experimentally that the alignment fluctuations manifest themselves as the noise of the linear birefringence. 
In a magnetic field, the spin-alignment fluctuations, in contrast to those of spin orientation, show up as the noise of the probe-beam ellipticity at the double Larmor frequency, with the most efficient geometry of its observation being the Faraday configuration with the light propagating along the magnetic field. 
We have detected the spin-alignment noise in a cesium-vapor cell probed at the wavelength of D2 line ($852.35$~nm). 
The magnetic-field and polarization dependence of the ellipticity noise are in full agreement with the developed theory.
\end{abstract}

\maketitle

\section{Introduction}
Properties of materials in physics are studied most frequently by measuring their response to external perturbations. 
In these measurements, the perturbation needed to reliably detect the response may noticeably affect characteristics of the studied system, thus hindering interpretation of the experimental data.  
For this reason, the signals that can be read out from the medium, in the absence of any external perturbation, in certain cases, appear to be highly valuable.  
The signals of this kind are usually revealed as spontaneous fluctuations of a certain physical quantity: electric current, pressure, magnetization, etc., with the information about the system contained in the spectra of these fluctuations. 
Investigation of these spectra is the subject of the \emph{noise spectroscopy}~\cite{Noise}.

This paper is devoted to one specific  type of this spectroscopy---to the spectroscopy of polarization noise of light caused by spin fluctuations of the medium probed by this light. 
Of particular value is the \emph{spin noise spectroscopy} (SNS) initially realized on atomic systems~\cite{Zap,Polz,Mitsui,Crook}, where the magnetization fluctuations are detected via the noise of the Faraday rotation. 
Nowadays, this technique provides indispensable information on the spin dynamics also useful for the development of spintronics.  
During the last years, this technique has revealed a number of unique capabilities~\cite{Oest2,Zap1,rise,Sin}. 
Specifically, the SNS made it possible to observe resonant magnetic susceptibility of nanosystem (quantum wells, quantum dots), inaccessible for the conventional EPR spectroscopy~\cite{Glazov,singlehole}, to examine magnetization dynamics of nuclei \cite{R,R1}, and to investigate nonlinear phenomena in such systems~\cite{Chal,Horn,R2,glazov:book}.  
The SNS technique  also allows one to distinguish between homogeneously and inhomogeneously broadened  lines of optical transitions~\cite{Zap2,PRA} and to measure homogeneous linewidth in an inhomogeneously broadened system using multi-probe noise technique~\cite{Crook1}; intensity modulation of the probe beam made it possible to expand the frequency range of signals detected in SNS up to microwave frequencies~\cite{Star,To1}. 
With the use of tightly focused light beams, the SNS allows one to investigate noise signals with high spatial resolution and even to realize tomography of magnetic properties of bulk materials~\cite{To}. 
The two-beam version of SNS proposed in~\cite{Koz} makes it possible, in principle, to observe both temporal and spatial correlations in magnetization. 

In some cases, studies of the spin noise appear to be in demand of ultrasensitive metrological measurements where a combination of the spin and alignment noise was detected for an atomic cloud in an optical dipole trap~\cite{ref3} and the spin alignment-to-orientation conversion was characterized~\cite{ref4}.
The prospects of application of electron and nuclear spins for quantum information processing and metrology~\cite{RB:add:1,RB:add:2,RB:add:3,RB:add:4,RB:add:5} make SNS particularly valuable for non- or weakly perturbative spin orientation detection. 
As opposed to spin-$1/2$ qubits, higher spin qutrits and qudits offer higher flexibility and substantial advantages for specific applications~\cite{high:1,high:2,high:3,high:4,soltamov:nat}.
In systems with spin larger than $1/2$ there are higher-order spin arrangements which are not reduced to the spin polarization. 
A simplest nontrivial one is the spin alignment where the average spin orientation is absent, but the spins are predominantly aligned parallel or antiparallel to a certain axis. 
Thus, in order to achieve full quantum control of the high-spin state~\cite{RB:1}, the perturbation-free techniques for  detection of higher-order spin arrangements and, particularly, alignment, need to be developed.

Conventional SNS is based on the detection of the spin fluctuations via fluctuations of \emph{gyrotropy} of the medium. 
In the presence of an external magnetic field typically applied in the Voigt geometry, the random fluctuations of spins precess with the Larmor frequency $\omega_L$ around the field. 
This Larmor precession is manifested as a peak at the frequency $\omega_L$ in the spin noise spectrum.

In this work, we show, both theoretically and experimentally, that for the spin larger than $1/2$ the spin-alignment fluctuations give rise to the noise of the \emph{linear birefringence} of the medium, which is manifested in the noise of ellipticity of the light transmitted through the medium. 
Depending on the experimental geometry, the spin-alignment fluctuations are revealed not only at the single frequency $\omega_L$, but also at the double frequency $2\omega_L$ allowing for unambiguous identification of the alignment noise.
Our experimental data on cesium vapors are in full agreement with the developed microscopic theory. 
These findings provide a stochastic counterpart of the effects of regular spin-precession-related oscillations of linear birefringence in magneto-optics~\cite{ref1,DoubleLarm,RB:add:7,ref2}.

\begin{figure}[t]
\subfigure{\label{fig1_a}}
\subfigure{\label{fig1_b}}
\subfigure{\label{fig1_c}}
\includegraphics[clip,width=.95\columnwidth]{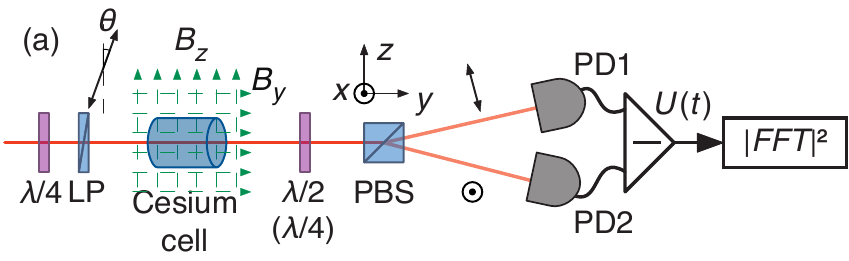}\\
\includegraphics[clip,width=.95\columnwidth]{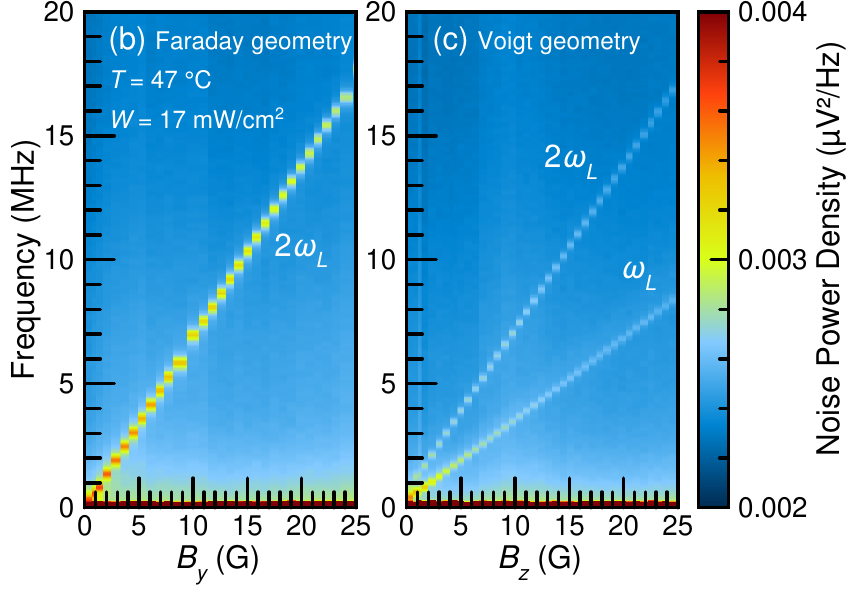}
\caption{(a) Schematics of the experimental setup. Magnetic-field dependence of the ellipticity-noise power spectrum of Cs in the Faraday (b) and Voigt (c) geometries. In the Voigt geometry, one can see the spin-noise peaks at the Larmor and double-Larmor frequencies, while in the Faraday geometry, the only peak at the double Larmor frequency is observed.}
\label{fig1}
\end{figure}

\section{Experimental results}
The experimental setup shown in Fig.~\ref{fig1_a} was modified as compared with the conventional arrangement used in the SNS. 
As a light source, we used a tunable single-frequency Ti:sapphire laser with ring resonator  providing the emission linewidth of $2$--$3$~kHz. 
The probe beam with a power of a few tens of mW was tuned approximately to the center of the long-wavelength hyperfine component of  the D2 line (where the spin-noise signal was more pronounced) and was expanded up to $\sim$10\ mm in diameter to minimize the effects of optical perturbation.    
The magnetic field $\bm B$ acting upon the cell with cesium  was created by a coil 20~cm in diameter that could be aligned  either along ($\bm B \parallel y$), or across the light beam propagation ($\bm B \parallel z$), thus providing either the Faraday or the Voigt geometry, respectively. 
Here, $y$ is the light propagation axis, $x$ and $z$ are the axes in the plane. 
The azimuth of the probe beam polarization could be varied by rotating linear polarizer (LP) in the beam circularly polarized with the aid of a quarter-wave plate ($\lambda/4$).  
After passing through the cesium cell, the probe beam hits the polarimetric detector comprised of a phase plate ($\lambda/2$ or $\lambda/4$), a polarization beamsplitter (PBS), and a differential photodetector. 
The half-wave plate was used when measuring the Faraday-rotation noise to balance the detector. 
To measure the ellipticity noise, the $\lambda/2$ plate was replaced by the $\lambda/4$ plate aligned with its axes at $45^\circ$ to polarizing directions of the PBS. 
In this case, the output signal of the differential photodetector vanishes when the input light beam is \emph{linearly polarized}, regardless of the azimuth of its polarization plane, and becomes nonzero only for \emph{nonzero ellipticity} of the input light.
The output signal of the differential detector was then fed to the FFT spectrum analyzer to obtain, after multiple averaging (during about 30 s), the power spectrum of the Faraday rotation or ellipticity noise. 
The intensity (light power density) of the probe beam in the cell was varied by changing the total light power and the beam cross section.

The cell with metal cesium was $50$~mm long and $30$~mm in diameter, without any antirelaxation coating, and did not contain any buffer gas.
The measurements of the polarization noise were performed at the wavelength of the probe beam tuned in resonance with the transition $6S_{1/2}\,(F = 4)\rightarrow 6P_{3/2}$ of the D2 line of Cs atoms ($\lambda =852.35$~nm), with the light intensity lying in the range $W = 10$--$100$~mW/cm$^2$. 
The density of cesium vapors could be varied by changing the temperature of the cell (in the range $30$--$80$~$^\circ$C) with the aid of a heat~gun.

The fluctuation spectra of the Faraday rotation contained, in the Voigt geometry, a single peak, with its  frequency linearly varied with the magnetic field ($\omega_L \propto B$). 
In the Faraday geometry, the Faraday rotation noise spectrum was concentrated in the vicinity of zero frequencies (see, e.g., Ref.~\cite{Zap1}). 
The spectral width of the Faraday rotation noise was mainly related, in our conditions, to inhomogeneity of the field across the sample.

\begin{figure}[t]
\includegraphics[width=.95\columnwidth,clip]{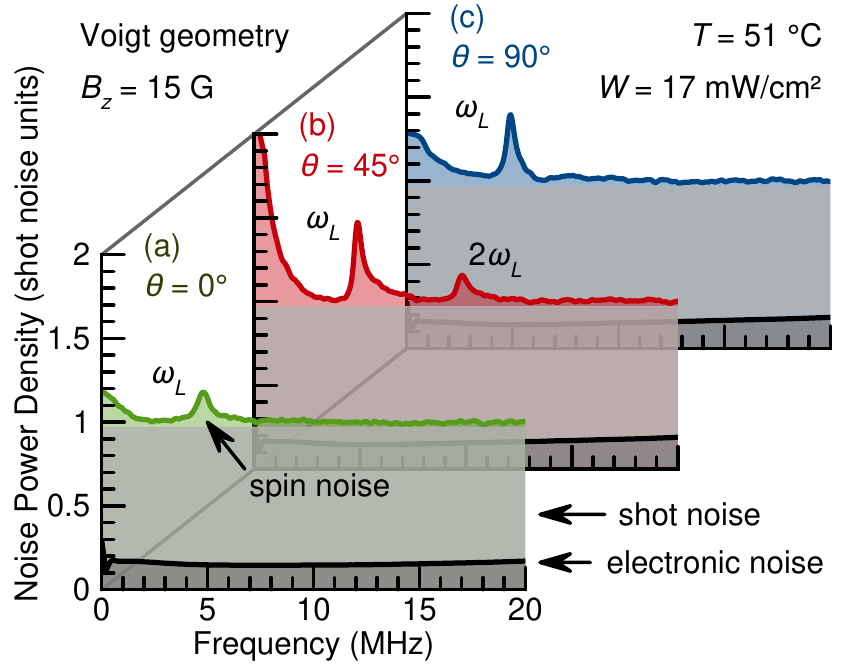}
\caption{Ellipticity noise spectra (in units of shot-noise power) detected at different angles $\theta$ between the probe beam polarization plane and magnetic field. The second harmonic of the Larmor frequency (at $\sim$$10$~MHz) is seen to be well pronounced at $\theta=45^\circ$ [curve (b)] and is not observed at $\theta= 0^\circ$ and $90^\circ$ [curves (a) and (c)]. The probe beam intensity is $17$~mW/cm$^2$. The spectra are normalized to the frequency response function of the detection channel.} 
\label{fig2}
\end{figure}

Key findings of our study are presented in Figs.~\ref{fig1_b} and \ref{fig1_c} where the \emph{ellipticity} noise is shown both in the Faraday and Voigt geometries. 
Strikingly, the ellipticity noise in the Faraday geometry, in addition to the zero-frequency peak, contains the peak at the double Larmor frequency, $2\omega_L$. 
In the Voigt geometry, in addition to the zero-frequency contribution, two peaks are seen at the single ($\omega_L$) and double  ($2\omega_L$) Larmor frequencies. 
Interestingly, in this geometry,  the intensity of the peak at the double Larmor frequency, $2\omega_L$, exhibits a strong dependence on the angle $\theta$ between the incident probe beam polarization vector and the magnetic field, reaching the highest value at $\theta=45^\circ$ and vanishing at $\theta=0^\circ$ and $90^\circ$, as illustrated in Fig.~\ref{fig2}. 
In other words, the double frequency component of the ellipticity noise vanishes if the probe beam is polarized either along or across the external magnetic field.

\begin{figure}[t]
\includegraphics[width=.95\columnwidth,clip]{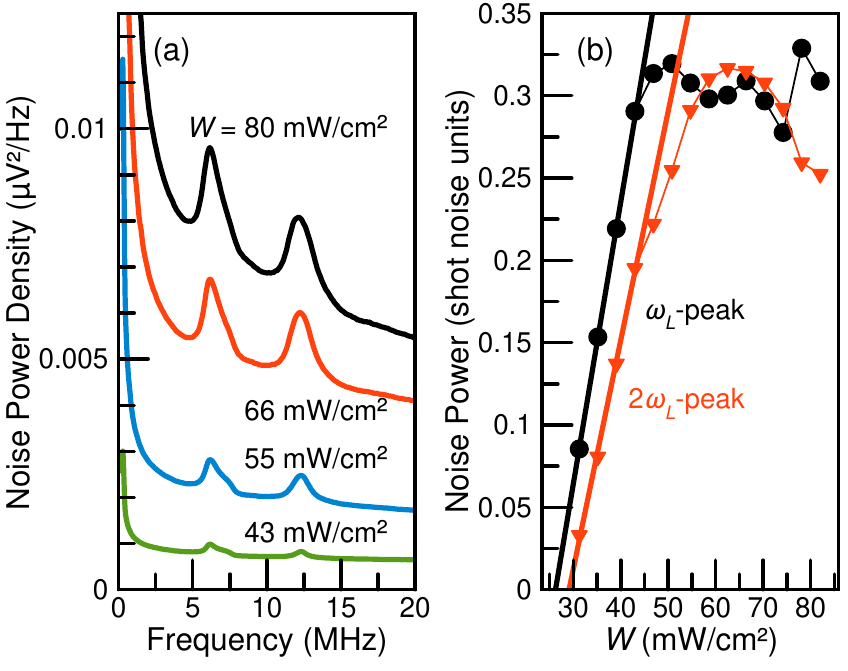}
\caption{The behavior of the $\omega_L$- and 2$\omega_L$-peaks of the ellipticity noise spectra of cesium at different values of the probe-beam intensity (a) and intensity dependence of the peak amplitudes (b) at $\theta=45^\circ$. The nonlinear growth of the shot noise power with increasing light intensity, clearly seen from the spectra in (a), indicates strong optical nonlinearity (bleaching) of cesium vapor at this level of light intensities. Still, the amplitudes of the two peaks, as seen from the dependence in (b), grow with the probe light intensity in a similar way. (The data in (a) are not normalized to the frequency response function of the detection system).} 
\label{fig3}
\end{figure}

The spectral width of the $\omega_L$ and $2\omega_L$ components of the ellipticity noise spectra is similar to that of the conventional spin noise contribution, suggesting that this broadening is also due to the inhomogeneity of the magnetic field. 
As seen from Fig.~\ref{fig3}, amplitudes of the two peaks in the ellipticity noise spectra grow in a similar way until bleaching of the cell is reached.  
This result shows that the peak at the double Larmor frequency is not related to effects of optical nonlinearity in cesium vapors.
\looseness=-1
 
From the viewpoint of symmetry, observation of the spin-precession signal (no matter, regular, or stochastic) in the Faraday geometry at the double Larmor frequency is natural.
Indeed, the spins oriented along the light beam  make the medium \emph{circularly anisotropic}, thus giving rise to the Faraday rotation.  
Oscillations of spin orientation with respect to the light propagation should evidently cause appropriate oscillation of the Faraday rotation.  
This is what we observe in the Voigt geometry of the conventional SNS. 
On the other hand, the spins \emph{aligned} along a certain direction across the light beam (not necessarily preferentially \emph{oriented}  along this direction) create a distinguished direction in the medium thus making it \emph{linearly anisotropic}. 
Rotation of the alignment direction around the light beam will evidently cause modulation of the light beam polarization at the double frequency of the rotation.  
This signal should be most conveniently observed  in the Faraday geometry. 
It is noteworthy that, in the Faraday geometry, the mean value of the birefringence is zero, which allows one to make measurements more correctly and to get rid of additional sources of noise associated with the static birefringence \cite{RB:add:3}. 
For the same reason, in the Voigt geometry, the contribution due to alignment oscillates at the $2\omega_L$ and also at $\omega_L$ for specific field orientations, see details in the model and discussion paragraphs below.
\looseness=-1
 
\begin{figure}[b]
\subfigure{\label{fig4_a}}
\subfigure{\label{fig4_b}}
\includegraphics[clip,width=.95\columnwidth, clip]{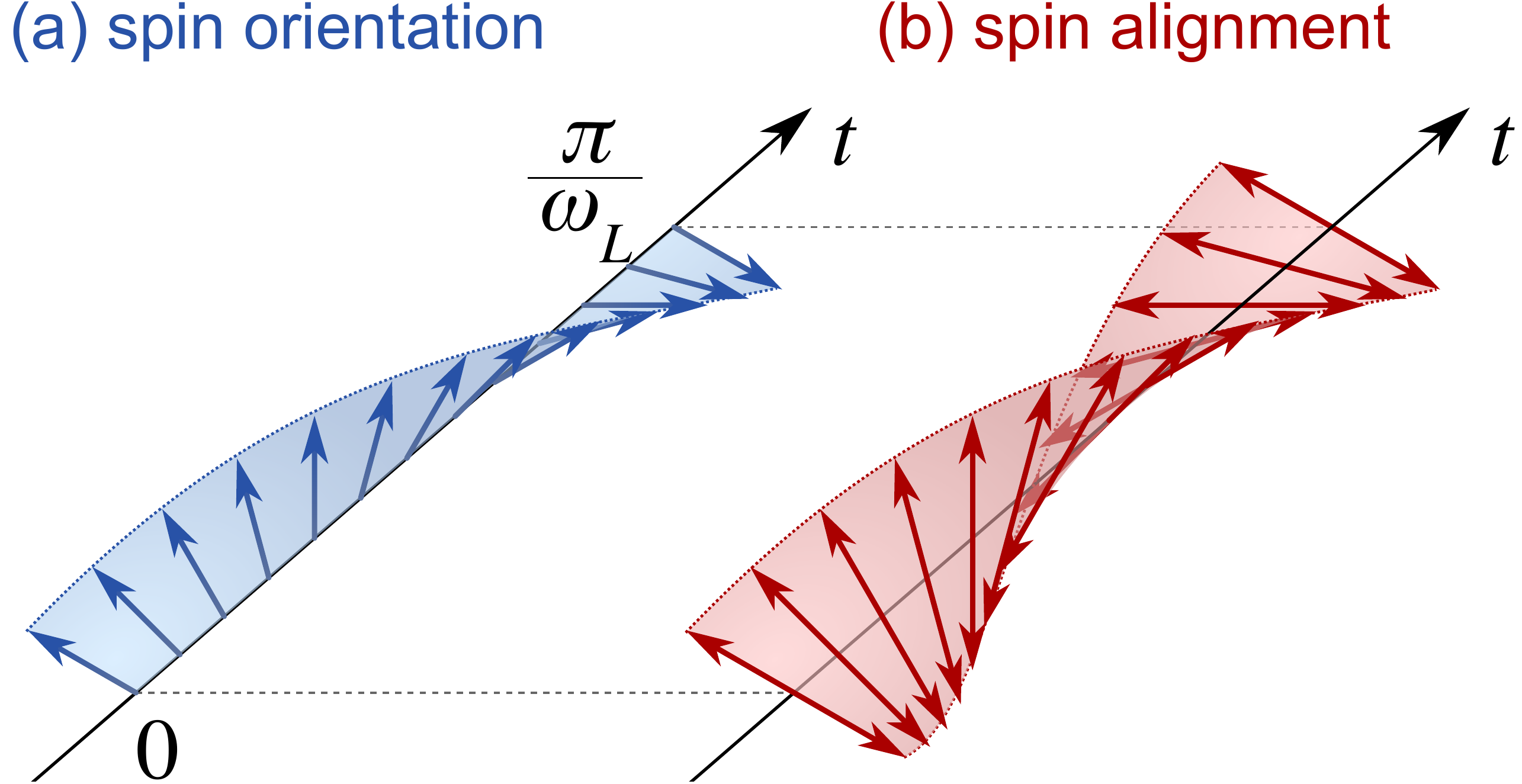}\\
\caption{Illustration of the precession effect on spin orientation (a) and spin alignment (b). 
The spin orientation after a half period of the precession appears to be inverted, while the spin alignment returns to its initial state.}
\label{fig4}
\end{figure}

Figure~\ref{fig4} illustrates schematically how the spin precession manifests itself in the precession of spin orientation [Fig.~\ref{fig4_a}] and spin alignment [Fig.~\ref{fig4_b}]. 
One can see that a half period of spin-orientation precession corresponds to the whole period of the alignment precession. 

Below we present a microscopic theory of the alignment fluctuations and discuss its manifestations in the ellipticity noise.

\section{Model and discussion} 
In conventional spin noise spectroscopy, the fluctuations of the Faraday or Kerr rotation are detected. 
In transparent isotropic media these fluctuations are related to the fluctuations of gyrotropy of the sample and can be phenomenologically associated with  the fluctuations of the asymmetric part of the dielectric susceptibility $\varepsilon_{\alpha\beta}$, $\alpha,\beta,\gamma=x,y,z$ are the Cartesian indices (cf.~\cite{RB:add:1,RB:add:2} for fully quantum description):
\begin{equation}
\label{depsilon:g}
\delta \varepsilon_{\alpha\beta} - \delta \varepsilon_{\beta\alpha} = \mathrm i A(\omega) \kappa_{\alpha\beta\gamma} \delta F_\gamma, 
\end{equation} 
where $\omega$ is the light frequency, $A(\omega)$ describes the efficiency of conversion of the spin fluctuation $\delta \bm F$ to the Faraday rotation, $\kappa_{\alpha\beta\gamma}$ is the Levi-Civita symbol. 
The analysis shows that for $F>1/2$ the fluctuations of a quadratic combination of spin components, i.e., fluctuations of spin alignment, contribute to the noise of the symmetric part of the dielectric susceptibility
\begin{equation}
\label{depsilon:b}
\delta \varepsilon_{\alpha\beta} + \delta \varepsilon_{\beta\alpha} = 2 S(\omega) \delta\{F_\alpha F_\beta\}_s, 
\end{equation}  
where $S(\omega)$ is the function of light frequency and $\{AB\}_s = (AB+BA)/2$ is the symmetrized product of the operators, giving rise to the birefringence noise. 
These fluctuations can be observed as the ellipticity noise. 
In the simplest model of a single homogeneously broadened resonance $A(\omega) = S(\omega) \equiv f /(\omega - \omega_0 + \mathrm i\gamma)$, where $\omega_0$ is the optical transition frequency, $\gamma$ is its homogeneous broadening and $f$ is proportional to the oscillator strength of the resonance.
\looseness=-1

Let us now calculate the autocorrelation function of the alignment fluctuations. 
In our experimental geometry, the quantity of interest is 
\begin{equation}
\label{corr}
\mathcal C_{zx}(\tau) = \langle \{F_z F_x\}_s(t+\tau) \{F_z F_x\}_s(t)\rangle, 
\end{equation}
and its power noise spectrum given by Fourier transform over $\tau$: $(\mathcal C_{zx})^2_\Omega = \int_{-\infty}^\infty \mathcal C_{zx}(\tau) \exp{(\mathrm i \Omega \tau)} d\tau$~\cite{glazov:book}. 
The averaging in Eq.~\eqref{corr} is carried out over the ensemble of atoms and over $t$ at a fixed $\tau$.  
For the Faraday geometry the spin precession in magnetic field couples the correlator $\mathcal C_{zx}$ in Eq.~\eqref{corr} with another correlator~(cf. Ref.~\cite{RB:2})
\begin{equation}
\label{C2}
\mathcal C_2(\tau) = \frac{1}{2} \langle \left[F_z^2(t+\tau) - F_x^2(t+\tau)\right] \{F_z F_x\}_s(t)\rangle.
\end{equation}  
Calculations show that the correlators in question obey the set of kinetic equations
\begin{subequations}
\label{dynamics:F}
\begin{align}
&\dot{\mathcal C}_{zx} = 2\omega_L \mathcal C_2 +\Gamma \mathcal C_{zx},\\
&\dot{\mathcal C}_{2} = -2\omega_L \mathcal C_{zx} +\Gamma \mathcal C_2,
\end{align}
\end{subequations}
where the dot on top denotes the time derivative, $\omega_L$ is the Larmor frequency and $\Gamma$ is the alignment relaxation rate. 
The initial conditions for Eqs.~\eqref{dynamics:F} can be immediately found from the equilibrium density matrix and, in the high-temperature limit, $k_B T \gg \hbar\omega_L$, read 
\begin{equation}
\label{Cxy:0}
C_{zx}(0) = \left[F(F+1)-\frac{3}{4}\right] \frac{F(F+1)}{15}, \quad \mathcal C_2(0)=0. 
\end{equation}
Hence, $C_{zx}(\tau) = C_{zx}(0) \cos{(2\omega_L \tau)} \exp{(-\Gamma|\tau|)}$ and the alignment noise power spectrum takes the form
\begin{equation}
\label{align:spec}
(\mathcal C_{zx})^2_\Omega = \pi \mathcal C_{zx}(0) \left[\Delta(\Omega - 2\omega_L) +\Delta(\Omega + 2\omega_L)\right],
\end{equation}
where we introduced the broadened $\delta$~function, $\Delta(\omega) = \Gamma/[\pi(\omega^2+\Gamma^2)]$. 
For the inhomogeneous magnetic field, Eq.~\eqref{align:spec} should be averaged over the values of $\omega_L$, which will give rise to an additional broadening of the peaks.

Thus the alignment fluctuations in the Faraday geometry occur at the double Larmor frequency. 
Indeed, while each of the spin components $F_z$ and $F_x$ precesses around the magnetic field with the single frequency $\omega_L$ their symmetrized product $\{F_zF_x\}_s$ rotates twice faster. 
It is confirmed by Eq.~\eqref{corr} and is in full agreement with the experimental data shown in Fig.~\ref{fig1_b}. 
In agreement with the general theory and qualitative expectations, the correlator $\mathcal C_{xz}$ vanishes for $F=1/2$.

The situation appears to be richer in the Voigt geometry. 
In order to analyze the fluctuations of alignment we derived a set of  six coupled differential equations for the correlators $\langle \{F_\alpha F_\beta\}_s(t+\tau) \{F_{\alpha'} F_{\beta'}\}_s(t)\rangle$ ($\alpha,\beta,\alpha',\beta'=x,y,z$) and the appropriate initial conditions. 
The solution in the time domain reads 
\begin{multline}
\label{solution}
\mathcal C_{zx}(\tau) = \mathcal C_{zx}(0) e^{-\Gamma |\tau|} \times \\
\left\{\frac{1}{4}\sin^2{2\theta}\left[ 3 + \cos{(2\omega_L \tau)}\right] + \cos^2{2\theta} \cos{(\omega_L \tau)} \right\},
\end{multline}
with $\mathcal C_{zx}(0)$ given by Eq.~\eqref{Cxy:0} and $\theta$ being the angle between the probe beam polarization and the magnetic field. 
Accordingly, the alignment noise power spectrum in the Voigt geometry reads
\begin{multline}
\label{align:spec:V}
(\mathcal C_{zx})^2_\Omega = \pi \mathcal C_{zx}(0) \Bigl\{\cos^2{2\theta}\left[\Delta(\Omega - \omega_L) +\Delta(\Omega + \omega_L)\right]\\
+ \frac{\sin^2{2\theta}}{4}\left[3\Delta(\Omega) +\Delta(\Omega - 2\omega_L) +\Delta(\Omega + 2\omega_L)\right] \Bigr\}.
\end{multline}

Here, the spectrum contains three components at $\Omega=0$, $\pm \omega_L$ and $\pm 2\omega_L$.\ 
The intensities of the components depend on the orientation of the probe polarization with respect to the magnetic field. 
At $\theta=0$ or $\pi/2$, only~the first harmonic of the Larmor frequency is present. 
Indeed, in this geometry, one of the factors in the product $\{F_zF_x\}_s$ is constant, while the other factor oscillates due to the spin precession around the field.
By contrast, at $\theta=\pi/4$ and $3\pi/4$, both $F_z$ and $F_x$ oscillate at the Larmor frequency as a function of time and their symmetrized product contains both the static contribution $\Omega=0$ and the second harmonic $\Omega=2\omega_L$.\ 
Again, our model analysis is fully in line with the experimental data shown in Figs.\ \ref{fig1_c} and \ref{fig2}.
Noteworthy, the presence of the peak at $\omega_L$ for $\theta=45^\circ$, Fig.\ \ref{fig2} is related to the fact that, in our experiments, the ellipticity noise is measured at small detunings with noticeable absorption. 
In this case, the spin fluctuations contributing to the gyrotropy noise [Eq.\ \eqref{depsilon:g}] observed at the single Larmor frequency manifest themselves in the ellipticity in addition to the alignment noise~\cite{note:ratio}.
Such spin fluctuations contribution to the ellipticity noise has been previously studied in various semiconductor systems~\cite{Glazov,glazov:book,Crook1}. 
The comparison of the widths of the $\omega_L$ and $2\omega_L$ peaks confirms that their broadening mainly originates from the field inhomogeniety.
\looseness=-1

Our experiments have shown that the $2\omega_L$ noise contribution vanished in the presence of a small amount of a buffer gas ($\sim$1 Torr of Ne). 
While the precise origin of the effect is not so far fully clear, one possible reason for this could be homogenization of the Doppler broadening which results in suppression of the birefringence fluctuations at small detunings~\cite{Zap2,RB:2}. 
Additionally, collisions with buffer gas can give rise to up to three times faster decay rates of alignment as compared with orientation are possible~\cite{dp:72,note:rel,az}. 
The inhomogeneity of the magnetic field is more pronounced for the $2\omega_L$ peak as well. Finally, non-linear effects cannot be ruled out completely.
\vspace{-1em}

\section{Conclusion} 
\vspace{-.3em}
We have shown that the spin noise spectra, usually detected via polarization noise of a probe laser beam, may reveal not only fluctuations of gyrotropy of the medium at Larmor frequency, but also fluctuations of linear birefringence at the double Larmor frequency.   
Physically, these two types of fluctuations are associated with random oscillations of spin polarization and spin alignment, respectively.  
According to our calculations and our experimental results, characteristics of the peak at the double Larmor frequency substantially differ from those of the conventional peak at Larmor frequency. 
Specifically, the intensity of the second-harmonic component is controlled by the orientation of the probe beam azimuth with respect to the magnetic field. 
Also, the second-harmonic peak appears to be more sensitive to the probe beam wavelength (being most pronounced inside the linewidth of the optical transition) and to conditions of spin motion (particularly, to the buffer gas pressure). 
At the same time, the alignment noise (as well as the noise of orientation) proves to be persistent at high light intensities, when optical nonlinearities cannot be ignored. Note that, in principle, under conditions of resonant probing, the effects of optical perturbation may become essential.   

Here, we used cesium vapors as a model system, however, the possibilities to observe and study the alignment noise by optical means go far beyond the field of atomic physics. 
This technique can be applied to any system with spins larger than $1/2$, including heavy-holes and excitons in semiconductors and nanosystems, spins of magnetic ions as well as the host lattice nuclei in solids paving way to practically perturbation-free detection of high spin states.
\vfill

\begin{acknowledgments}
\vspace{-.6em}
This work was supported by the Russian Science Foundation (grant No.\ 17-12-01124). 
The theoretical contribution of M.M.G. was supported by SPSU research grant ID\ 40847559. 
I.I.R. acknowledges President of the Russian Federation Grant No.\ MK-2070.2018.2 for support of the preliminary spectroscopic characterization of the cesium vapor cells.
The work was performed using equipment of the SPbU Resource Center ``Nanophotonics''.
\end{acknowledgments}


\begin{thebibliography}{}
\bibitem{note:rel} The spin relaxation is caused by fluctuating magnetic fields, either real or induced by the spin-orbit interaction. The alignment relaxation can be caused, e.g., by the fluctuating electric field gradients. Such perturbations are even at the time reversal as compared with those causing the spin relaxation.
\end{thebibliography}
\end{document}